\documentclass[epsf,11pt]{article}
\usepackage{epsfig,graphicx}

\catcode`@=11 
\def\lsim{\mathrel{\mathpalette\@versim<}}
\def\gsim{\mathrel{\mathpalette\@versim>}}
 \def\@versim#1#2{\lower0.2ex\vbox{\baselineskip\z@skip\lineskip\z@skip
       \lineskiplimit\z@\ialign{$\m@th#1\hfil##$\crcr#2\crcr\sim\crcr}}}
\catcode`@=12 

\begin{document}

\pagestyle{empty}
{\large
\begin{center}

{\Large\bf How Resummation Depresses the Gluon at Small $x$ }\\
\bigskip
\bigskip
{\large\bf  Stefano Forte}\\
\smallskip
{ \it Dipartimento di Fisica, Universit\`a di Milano and\\ 
INFN, Sezione di Milano,\\ Via Celoria 16, I-20133 Milano, Italy}\\
\bigskip
{\large\bf  Guido Altarelli}\\
\smallskip
{ \it CERN, Department of Physics, Theory Division,\\
CH--1211 Gen\`eve 23, Switzwerland,\\
Dipartimento di  Fisica``E.Amaldi'', 
Universit\`a Roma Tre\\
 INFN, Sezione di Roma Tre\\ via della Vasca Navale 84, I--00146 Roma, Italy}\\
\bigskip
{\large\bf  Richard D. Ball}\\

\smallskip
{ \it School of Physics, University of Edinburgh\\Edinburgh EH9 3JZ,
 Scotland}

\vskip.4cm

{\bf Abstract}\\
\end{center}
\noindent
We summarize recent progress in the resummation of
  perturbative evolution at small $x$. We show that the problem of
  incorporating BFKL small $x$ logs in GLAP evolution is now
  completely solved, and that the main  effect of small $x$ resummation is
  to reduce the growth of the gluon at small $x$ in the HERA and LHC regions.\hfill\\
\vskip.3cm
\begin{center}
 Talk at \\
{\bf DIS 2006}\\
Tsukuba, Japan, April  2006\\
{\it to be published in the proceedings}
\end{center}
\vfill
June 2006\hfill IFUM-867/FT}
\eject
\setcounter{page}{1} \pagestyle{plain}
\section{BFKL logs in GLAP evolution}
It has been thought for a long time that the resummation of small $x$ 
contributions to perturbative evolution
might require techniques that go beyond standard perturbative
factorization. However, dramatic theoretical progress,\cite{Forte:2005mw,Salam:2004rh} largely
prompted by the unexpected success of NLO GLAP evolution 
in describing the growth of structure functions observed at
HERA,\cite{Ball:1994du,Dittmar:2005ed} has led to a complete understanding of
this issue within a perturbative framework. Namely, it is now clear that
small $x$ logs, as described by the BFKL equation, can be fully
incorporated in the standard GLAP framework, that their inclusion
stabilizes the behaviour of perturbation theory at small $x$, and that, somewhat
surprizingly, they suppress the growth of parton distributions at
small $x$ down to the smallest values of $x$ and the largest values of
$Q^2$ accessible at HERA and the LHC.
\section{Theoretical progress: the three ingredients}
The resummation of logarithmically enhanced small $x$
contributions to perturbative evolution can be performed within two
different approaches, which share several basic physical assumptions
but differ in the implementation, most notably because one
(CCSS\cite{Ciafaloni:1999yw,Ciafaloni:2003rd}) 
is rooted in the  BFKL equation and extracts the
anomalous dimension numerically from the gluon green function, while
the other (ABF,\cite{Altarelli:1999vw,Altarelli:2001ji,Altarelli:2005ni}
on which we will concentrate) improves the standard GLAP anomalous
dimension by including in it infinite series of logarithmically enhanced
terms. A detailed comparison is in
Ref.\cite{Dittmar:2005ed}.

\begin{figure}[ht]
\centerline{
\includegraphics[width=.43\linewidth]{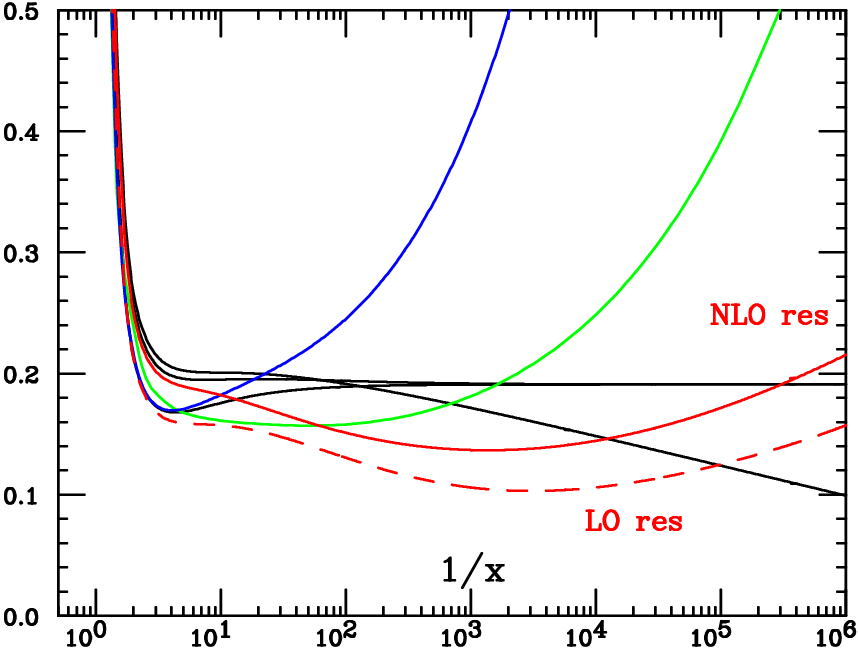}
\includegraphics[width=.47\linewidth]{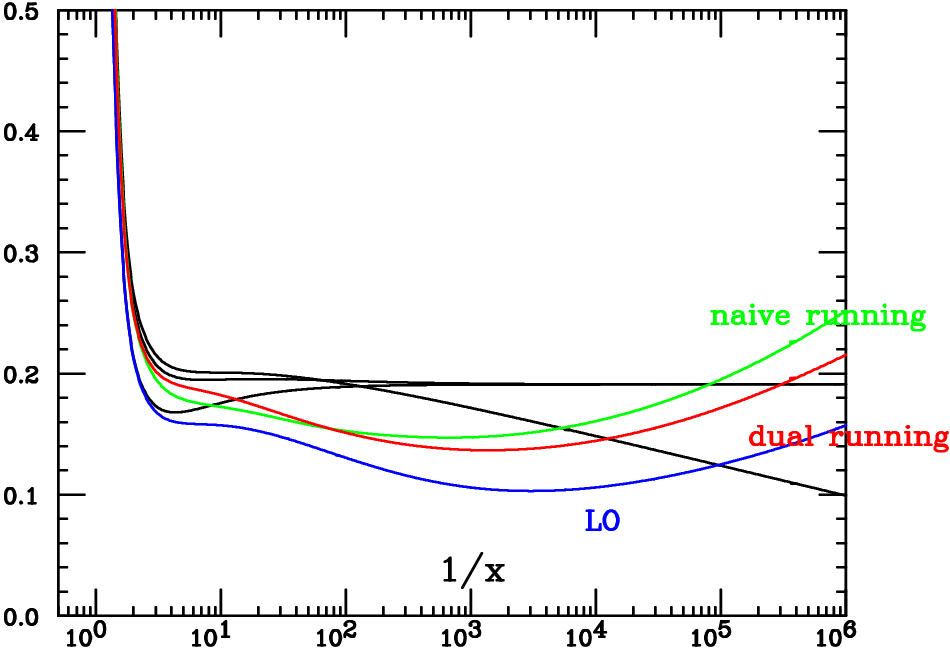}}
\caption{(Left) Resummed and unresummed splitting functions with
  $\alpha_s=0.2$ and $n_f=0$. The
curves shown are (from top to bottom on the right): double--leading
resummation, running coupling resummation thereof, full NLO
resummation, fixed LO and NLO, full LO
resummation, fixed NNLO. (Right) Effect of the choice of argument of the running coupling. }
\end{figure}
The ABF resummation is based on three basic ingredients.
The first ingredient is  duality (first introduced in
Ref.\cite{Ball:1997vf,Altarelli:1999vw} and recently\cite{Ball:2005mj} proven to all
orders at the running--coupling level), which states that at leading
twist the BFKL and GLAP equations admit the same solutions if their
respective kernels are suitably matched. Using duality, the
information on leading, subleading,\dots $x$ logs from BFKL and that
on leading,
subleading,\dots $Q^2$ logs from GLAP can be combined in a
double--leading expansion of either (or both) the GLAP and BFKL
kernels. The double--leading splitting function thus reabsorbs the fixed--order
small--$x$ instability, but it leads to a splitting function (see
Fig.~1) which grows too rapidly at small $x$ in comparison to the
data,\cite{Altarelli:2000mh} thereby signalling the need for further
resummation. 

The second ingredient is the all--order resummation of
small $x$ running coupling effects: these are formally subleading, but
their contribution to the splitting function diverges as $x\to
0$. Their resummation\cite{Altarelli:2001ji} qualitatively changes
the small $N$ behaviour of the anomalous dimension: the double--leading square--root
cut at relatively  large $N\sim 0.5$ (at the HERA scale)  
is replaced by a simple pole with very
small residue at small  $N\sim 0.2$. This considerably softens the
small $x$ behaviour of the splitting function, as shown in
Fig.~1. 

Finally, the third ingredient is the symmetry of the BFKL
kernel,\cite{Salam:1998tj} which relates the collinear GLAP region to
an anti-collinear region, where incoming and outgoing gluon
virtualities are exchanged. If combined with the running coupling
resummation,\cite{Altarelli:2005ni} this symmetry further softens the
small $x$ behaviour and, more importantly, it leads to a stable
perturbative expansion at the resummed level. 
The interplay of this symmetry with running
coupling effects is quite subtle, because the running of the coupling
breaks the collinear--anticollinear symmetry. In Fig.~1 the
NLO resummed splitting function is compared with that where the
argument of the strong coupling is chosen ``naively'' (i.e. just $Q^2$)
instead of that obtained from running coupling duality: the effect is 
almost as large as the whole NLO correction.

\section{Resummed evolution: stability and softening}

The resummation has the effect of completely stabilizing the splitting
function at small $x$''\cite{Altarelli:2005ni} the NLO and LO resummed
results
are quite close, they all depend weakly on the value of $\alpha_s$ and
on the choice of factorization scale, and resummed evolution is not
very different from unresummed one, thereby explaining the success of standard
GLAP evolution at HERA.

In fact, 
the qualitative behaviour of fully resummed results completely
differs from standard BFKL folklore, according to which 
leading small $x$ logs should lead to a strong
growth of the structure function at small $x$, while subleading
logs  change this behaviour completely, by replacing it with
a softer growth (or possibly no growth at all).  Quite on the
contrary, it is clear from Figure~1 that the resummed splitting
function lies below the unresummed 
down to very small $x\lsim 10^{-5}$, and that the NLO resummed result
is similar to the LO one, but in fact somewhat above it.

\begin{figure}[ht]
\centerline{
\includegraphics[width=.3\linewidth]{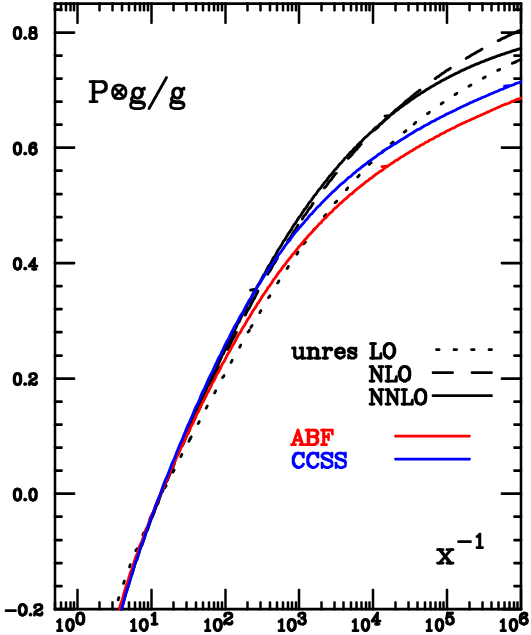}
\includegraphics[width=.29\linewidth]{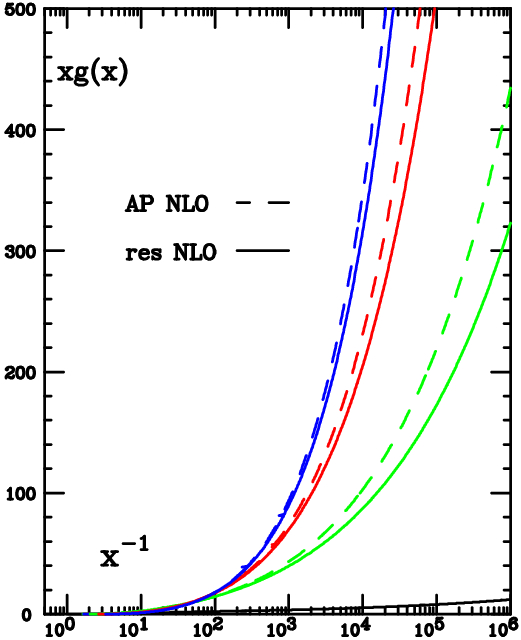}
\includegraphics[width=.38\linewidth]{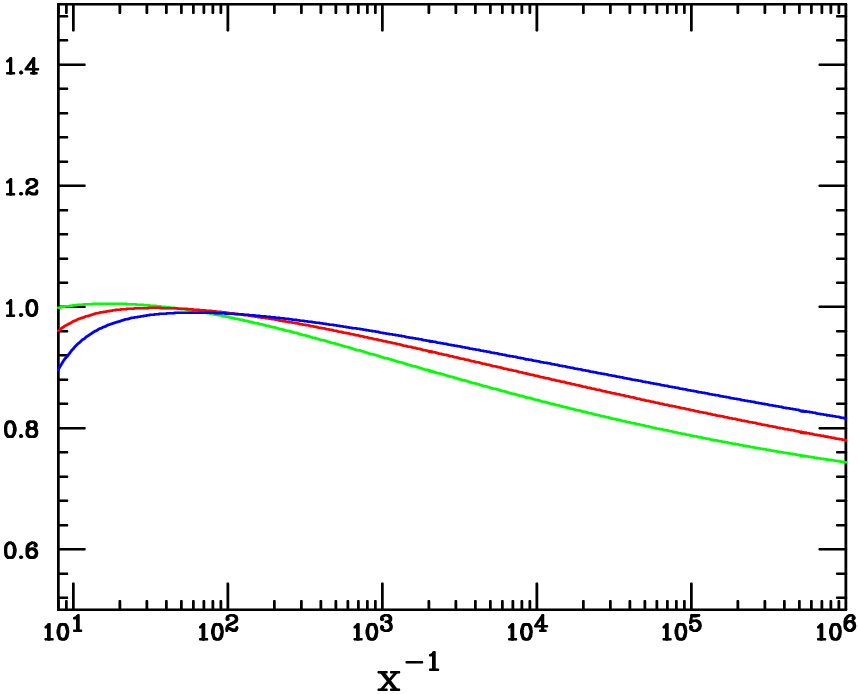}}
\caption{(left) Log derivative of the gluon with respect to scale at $Q^2=4$~GeV$^2$; (center) 
  gluon evolution (from bottom to top: $Q^2=$~4, $10^2$, $10^4$,
  $10^6$~GeV$^2$); 
(right) resummed/fixed NLO $K$--factor (from bottom to top: $Q^2=10^2$, $10^4$,
  $10^6$~GeV$^2$) .}
\end{figure}
These conclusions are particularly evident if one considers the
impact of resummation (in the $n_f=0$ case) on the evolution of the
gluon distribution (Figure~2). The resummed result  for
the convolution of the
splitting function with a gluon distribution of the form
$xg(x,4\hbox{GeV}^2)=x^{-0.18} (1-x)^5$   is below the unresummed LO, NLO,
and NNLO results down to very small $x$. The effect of this suppression
of the scale dependence is moderate but visible in the evolution of
the gluon up to rather large values of $Q^2$ and $1/x$, also shown.   The
corresponding $K$--factor (defined as the ratio of the NLO resummed to
NLO fixed order evolved gluon) depends rather weakly on scale, as the
figure shows. 

\section{Outlook}
Resummed results at small $x$ show remarkable stability, and suppress
the growth of parton distributions in a moderate but visible way. It
will be interesting to work out their phenomenological implications,
both at the LHC and in other relevant context such as high--energy
cosmic rays.

\end{document}